\documentclass[a4paper,twocolumn]{article}
\usepackage{spconf}
\usepackage{amsfonts}
\usepackage{graphicx}
\usepackage{amsmath}

\title{New indices of coherence for one and two-dimensional fields}
\name{Bernard Lacaze}
\address{TESA Laboratory\\
	Signal Processing Unit\\
	7 Boulevard de la Gare - 31500 TOULOUSE - FRANCE\\
    bernard.lacaze@tesa.prd.fr
}

\begin{document}

\maketitle

\begin{abstract}
The modern definition of optical coherence highlights a frequency dependent
function based on a matrix of spectra and cross-spectra. Due to general
properties of matrices, such a function is invariant in changes of basis. In
this article, we attempt to measure the proximity of two stationary fields
by a real and positive number between 0 and 1. The extremal values will
correspond to uncorrelation and linear dependence, similar to a correlation
coefficient which measures linear links between two random variables. We
show that these "indices of coherence" are generally not symmetric, and not
unique. We study and we illustrate this problem together for one-dimensional
and two-dimensional fields in the framework of stationary processes.

\textit{keywords: }Coherence, optical beams, stationary processes, linear
invariant filters.
\end{abstract}

\section{Introduction}

Originally, the coherence of an optical beam measured its ability to
interfere. The beam can be modelled by a "quasi-monochromatic"
one-dimensional process, which means that the power spectrum is close to a
line at some frequency $\omega _{0}/2\pi .$ We know that a thin spectral
line allows a larger number of franges than a line with a larger width. The
spectral width is related to the auto-correlation function which decreases
faster when the width increases. In this simple situation, the coherence $%
\gamma \left( \tau \right) $ can be defined as the reduced auto-correlation
function, and then takes the value $1$ for $\tau =0\,,$ and the value 0 for
large values of $\tau ,\,$\ except for the ideal monochromatic wave, which
is always an idealization.

When comparing two one-dimensional processes (the model is now
two-dimensional), the "complex degree of coherence" $\gamma _{12}\left( \tau
\right) $ is \ defined by \cite{Born}, \cite{Wolf2}%
\begin{equation}
\gamma _{12}\left( \tau \right) =\frac{K_{12}\left( \tau \right) }{\sqrt{%
K_{1}\left( 0\right) K_{2}\left( 0\right) }}
\end{equation}%
where $K_{1},K_{2},K_{12}$ are the auto-correlations and the
cross-correlation between both processes. We have $0\leq \left\vert \gamma
_{12}\left( \tau \right) \right\vert \leq 1$ by the Cauchy-Schwarz
inequality, but it is possible that $\gamma _{12}\left( \tau \right) $ does
not reach the value 1, and the maximal value has a particular sense. Optical
beams have a two-dimensional electrical field orthogonal to the trajectory.
The "mutual coherence" between two points $P_{1}$ and $P_{2}$ is defined
from the "electric mutual coherence matrix" of correlation and
cross-correlation functions \cite{Wolf}%
\begin{equation}
\mathbf{M}\left( \tau \right) =\left[
\begin{array}{cc}
K_{11}\left( \tau \right) & K_{12}\left( \tau \right) \\
K_{21}\left( \tau \right) & K_{22}\left( \tau \right)%
\end{array}%
\right]
\end{equation}%
where $K_{ij}\left( \tau \right) =$E$\left[ E_{i}\left( P_{1},t\right)
E_{j}^{\ast }\left( P_{2},t-\tau \right) \right] ,$ $E_{i}\left(
P_{k},t\right) $ being the component $i$ ($i=1,2)$ of the field at the point
$P_{k}$ ($k=1,2)$ in some orthogonal basis$.$ The "complex degree of
coherence" is defined by ($I_{1}$ and $I_{2}$ are the intensities) \cite%
{Wolf2}%
\begin{equation}
\gamma \left( \tau \right) =\frac{\text{tr}\mathbf{M}\left( \tau \right) }{%
\sqrt{I_{1}I_{2}}}.
\end{equation}%
It is worth noting that this quantity does not depend on the orthogonal
basis of reference because tr$\mathbf{M}\left( \tau \right) ,I_{1},I_{2}$
are matricial invariants \cite{Terv1}.

Rather than working with correlations, modern optics consider spectral and
cross-spectral matrices \cite{Wolf}. The "spectral degree of coherence" has
the same shape as $\left( 3\right) $ except that it is a function of the
frequency $\omega /2\pi $ through the "electric cross-spectral density"$.$
Whatever the framework, the "degree" of coherence is a real or complex
quantity which may depend on the time or on the frequency but which may be
constant. It is the case in a number of domains of physical or human
sciences from astrophysics to demography. Here, its modulus takes the
extreme values 0 and 1 in very particular circumstances of linkage between
the processes taken into account.

Formulas $\left( 1\right) $ and $\left( 3\right) $ were fitted to
interferences plans. For values $\tau _{0}$ such that $\gamma _{12}\left(
\tau _{0}\right) =1,$ powers of processes are added for the delay $\tau
_{0}. $ Moreover, this means that both processes taken into account can be
deduced by a linear operation characterized by a Linear Invariant Filter
(LIF) (see the following section). Conversely, if $\left\vert \gamma
_{12}\left( \tau \right) \right\vert $ cannot take the value 1, this means
that some parts of both processes are uncorrelated, even if other parts are
very strongly linked. It is clear that the latter parts can lead to
interferences from matched devices, but not the first parts. Clearly, it is
very important to study this kind of decomposition, and to deduce measures
of "distances" between processes. We will deduce a reasonable family of
"indices of coherence".

Let assume that we look at two beams defined by three uncorrelated processes
\textbf{A, B }and \textbf{C. }The beams
\begin{equation}
\mathbf{D=A+B\ }\text{and }\mathbf{E=A+C}
\end{equation}%
\textbf{\ \ \ \ \ }can generate good plans of interference only when the
"intensities" of \textbf{B }and \textbf{C }are weak with respect to \textbf{%
A. }In such a decomposition, we see the parts which interfere (\textbf{A}
with itself\textbf{)} and the parts which cannot interfere (\textbf{B} with
\textbf{C} and \textbf{A }for instance\textbf{)}. Conversely, to provide
such a decomposition gives strong informations on the ability of beams to
interfere.

The notion of coherence has to be put in front of neighboorhood, proximity,
distance, common points between functions, random variables, random
processes, or family of random processes. In the framework developed here,
Hilbert spaces of random variables and linear algebra allow the simplest
theoretical developments.

In this article, we look for an "index of coherence" which expresses the
proximity of some processes. It will be a positive number between 0 and 1,
the extremal values being reserved to the uncorrelation and the total
dependence. The next section studies the one-dimensional case, where we look
for links between two one-dimensional processes. The third section provides
a generalization to two-dimensional random processes. Simple examples are
developed in both cases. Appendices recall used mathematical tools and too
long proofs.

\section{One-dimensional case}

\subsection{A family of indices of coherence}

1) Let \textbf{U, V} be\textbf{\ }two stationary random processes with
elements $U\left( t\right) ,V\left( t\right) ,t\in \mathbb{R},$ power
spectral densities $s_{U}\left( \omega \right) ,s_{V}\left( \omega \right) ,$
cross-spectrum $s_{UV}\left( \omega \right) $ (see appendix 1). We assume
that the supports of $s_{U}\left( \omega \right) $ and $s_{V}\left( \omega
\right) $ are identical.

Let consider the linear invariant filter (LIF) $\mathcal{F}$ with complex
gain $\phi \left( \omega \right) $ defined by (see appendix 1)%
\begin{equation}
\phi \left( \omega \right) =\left[ \frac{s_{VU}}{s_{U}}\right] \left( \omega
\right) .
\end{equation}%
When the processes \textbf{A }and \textbf{B} verify%
\begin{equation}
\left\{
\begin{array}{c}
A\left( t\right) =\mathcal{F}\left[ \mathbf{U}\right] \left( t\right) \\
V\left( t\right) =A\left( t\right) +B\left( t\right)%
\end{array}%
\right.
\end{equation}%
the processes \textbf{A }and \textbf{B} become orthogonal (E$\left[ A\left(
t\right) B^{\ast }\left( u\right) \right] =0)$ and
\begin{equation}
\left\{
\begin{array}{c}
A\left( t\right) \in \mathbf{H}_{U},B\left( t\right) \perp \mathbf{H}_{U} \\
\text{E}\left[ \left\vert V\left( t\right) \right\vert ^{2}\right] =\text{E}%
\left[ \left\vert A\left( t\right) \right\vert ^{2}\right] +\text{E}\left[
\left\vert B\left( t\right) \right\vert ^{2}\right]%
\end{array}%
\right.
\end{equation}%
where $\mathbf{H}_{U}$ is the Hilbert space of linear combinations of the $%
U\left( t\right) $ when $t$ spans $\mathbb{R}$ (see appendix 2)$.$

Formula $\left( 6\right) $ splits $V\left( t\right) $ into two parts, the
first one $A\left( t\right) $ which belongs to \textbf{H}$_{U}$ (it is a
linear combination of the $U\left( u\right) ,u\in \mathbb{R}$), and the
second one $B\left( t\right) $ which is orthogonal to \textbf{H}$_{U}$ (i.e.
uncorrelated with the $U\left( u\right) ,u\in \mathbb{R}).$

In decomposition $\left( 6\right) $, \textbf{V }and \textbf{U }are
"neighbouring" when $E\left[ \left\vert B\left( t\right) \right\vert ^{2}%
\right] $ is weak compared to $E\left[ \left\vert A\left( t\right)
\right\vert ^{2}\right] $, which is equivalent to a strong "coherence"
between them. Conversely, the "coherence" is weak when this ratio is too
large. In this context, the "distance" between \textbf{V }and \textbf{U }is
not defined by a accurate relation taking into account the r.v. $U\left(
t\right) $ and $V\left( t\right) ,$ but a "distance" between for instance $%
U\left( t\right) $ and \textbf{H}$_{V}$ (spanned by the set $V\left(
u\right) ,u\in \mathbb{R}$). These considerations allow to define "indices
of coherence" which are constants and not some functions of the time or the
frequency. \ Let consider $\rho _{UV}$ defined by%
\begin{equation}
\begin{array}{c}
\rho _{UV}=\frac{\text{E}\left[ \left\vert A\left( t\right) \right\vert ^{2}%
\right] }{\text{E}\left[ \left\vert V\left( t\right) \right\vert ^{2}\right]
}=\frac{1}{\sigma _{V}^{2}}\int_{-\infty }^{\infty }\left[ \frac{\left\vert
s_{VU}\right\vert ^{2}}{s_{U}}\right] \left( \omega \right) d\omega \\
\sigma _{V}^{2}=K_{V}\left( 0\right) =\int_{-\infty }^{\infty }s_{V}\left(
\omega \right) d\omega .%
\end{array}%
\end{equation}%
We have $\rho _{UV}\in \left[ 0,1\right] ,$ and%
\begin{equation}
\left\{
\begin{array}{c}
\rho _{UV}=1\Longleftrightarrow V\left( t\right) =\mathcal{F}\left[ \mathbf{U%
}\right] \left( t\right) \\
\rho _{UV}=0\Longleftrightarrow V\left( t\right) \perp \mathbf{H}_{U}.%
\end{array}%
\right.
\end{equation}%
In the first case, the process \textbf{V }can be (linearly) reconstructed
from \textbf{U,} and both processes \textbf{U }and \textbf{V }are orthogonal
in the second case. $A\left( t\right) $ is the part of $V\left( t\right) $
which is explained by \textbf{U }i.e. by the set of random variables $%
U\left( u\right) ,u\in \mathbb{R}$ together than $\rho _{UV}$ is a measure
of the part of the power of $V\left( t\right) $ which is explained by
\textbf{U}.

2) In the case $\rho _{UV}=1$ $\left( \mathbf{B=0}\right) $, both processes
are "coherent" (\textbf{U }defines completely \textbf{V }and conversely),
and "uncoherent" when $\rho _{UV}=0$ (\textbf{U }and \textbf{V }are
orthogonal). For $\rho _{UV}\neq 0$ and 1, they are "partially coherent". It
is useless to add the redundant terms "totally" or "completely" or other
qualifiers \cite{Laca4}, \cite{Laca2}. Clearly, $\rho _{UV}$ has the
qualities that we expect for an "index of coherence". \ The fact that $\rho
_{UV}$ is deduced from a perfectly defined orthogonal decomposition is a
strong supplementary argument. It is worth noting that, using $\left(
6\right) ,\left( 7\right) $%
\begin{equation}
\rho _{\mathcal{G}\left[ \mathbf{U}\right] V}=\rho _{UV}
\end{equation}%
where $\mathcal{G}$ is some LIF (with complex gain which does not cancel).
This equality shows that $\rho _{UV}$ measures the proximity of \textbf{H}$%
_{U}$ with $V\left( t\right) .$ We obtain the same value of the index of
coherence whatever the stationary process chosen in \textbf{H}$_{U}.$

Unfortunately, this index is not symmetric, except for the bounds 0 and 1.
We have%
\begin{equation*}
\left\{
\begin{array}{c}
\rho _{UV}-\rho _{VU}=\int_{-\infty }^{\infty }\left[ \frac{\left\vert
s_{VU}\right\vert 2}{s_{U}s_{V}}\upsilon \right] \left( \omega \right)
d\omega \\
\upsilon \left( \omega \right) =\left[ \frac{s_{V}}{\sigma _{V}^{2}}-\frac{%
s_{U}}{\sigma _{U}^{2}}\right] \left( \omega \right)%
\end{array}%
\right.
\end{equation*}%
This quantity has no reason to cancel, except if we add hypotheses, for
instance the equality of the normalized spectra. Moreover, the family of $%
\rho _{a}$ defined by%
\begin{equation}
\rho _{a}=a\rho _{UV}+\left( 1-a\right) \rho _{VU}
\end{equation}%
verifies the conditions above when $a\in \left[ 0,1\right] .$ Then, it is
easy to construct families of positive numbers which illustrate links
between two stationary processes. Obviously, $\rho _{1/2}$ is symmetric, and
it is the only one symmetric provided that $\rho _{UV}\neq \rho _{VU}.$

3) Now, let assume that the support of $s_{U}$ (and $s_{V})$ can be split in
the sets $\Delta $ and $\Delta ^{\prime }$ of positive measure such that
\begin{equation*}
\left\{
\begin{array}{c}
s_{UV}\left( \omega \right) =0,\omega \in \Delta \\
s_{UV}\left( \omega \right) \neq 0,\omega \in \Delta ^{\prime }.%
\end{array}%
\right.
\end{equation*}%
In decomposition $\left( 6\right) $ we have at the same time%
\begin{equation*}
\left\{
\begin{array}{c}
s_{A}\left( \omega \right) =\left[ \frac{\left\vert s_{VU}\right\vert ^{2}}{%
s_{U}}\right] \left( \omega \right) =0,\omega \in \Delta \\
s_{A}\left( \omega \right) >0,\omega \in \Delta ^{\prime }%
\end{array}%
\right.
\end{equation*}%
which implies $0<\rho _{UV}<1.$ By symmetry, it is the same for $\rho _{VU}.$
Obviously, we find the same result when the supports of $s_{U}$ and $s_{V}$
are distinct.

4) Let assume that \textbf{U }and \textbf{V} model an optical beam, where a
source, a direction and a sense of propagation are given. If \textbf{U }is
nearer the source than \textbf{V,} the decomposition $\left( 6\right) $ is
natural, because we can consider that \textbf{U }is a source for \textbf{V}.%
\textbf{\ }In the latter, we expect to find a component closely linked to
\textbf{U} added to a noise which models a loss of information. \textbf{A }%
and \textbf{B }answer this question. This point of view is accurately
expressed by $\left( 7\right) $ and by the index $\rho _{UV}$ of $\left(
8\right) $ rather than by $\rho _{VU}$ based on a decomposition of $U\left(
t\right) .$

Whatever the definition of the index of coherence, based on the
decomposition of one or both processes \textbf{U} and \textbf{V}, it is
clear that the decomposition itself in two processes (for instance \textbf{A}
and \textbf{B}) gives more insights about links between the processes than
any index only based on statistical moments. When the beam is not reduced to
a ray but fills some volume close to some axis, the relative places of the
processes are more difficult to characterize.

To summarize, we have defined a real and positive index $\rho _{UV}$ which
takes its values in $\left[ 0,1\right] $ and which expresses the proximity
between \textbf{U }and \textbf{V. }We will say that both processes are
"coherent" when $\rho _{UV}=1,$ and "uncoherent" when $\rho _{UV}=0$ (rather
than fully or completely coherent or uncoherent)$.$ In other cases, they
will be "partially coherent". Actually, through $\left( 11\right) ,$ we have
shown that $\rho _{UV}$ and $\rho _{VU}$ define a family of indices $\rho
_{a}$ ($0\leq a\leq 1)$ with the same properties, added to the symmetry
property for $\rho _{1/2}$.

\subsection{Estimation and index of coherence}

When the LIF $\mathcal{F}^{-1}$ is well defined, $\left( 6\right) $ is
equivalent to
\begin{equation*}
\left\{
\begin{array}{c}
W\left( t\right) =\mathcal{F}^{-1}\left[ \mathbf{V}\right] \left( t\right)
=U\left( t\right) +C\left( t\right) \\
C\left( t\right) =\mathcal{F}^{-1}\left[ \mathbf{B}\right] \left( t\right)
\\
U\left( t\right) =\mathcal{F}^{-1}\left[ \mathbf{A}\right] \left( t\right) .%
\end{array}%
\right.
\end{equation*}%
Both processes \textbf{U }and \textbf{C} are uncorrelated. We look for an
estimation of \textbf{U }from the observation of \textbf{V }(equivalently
from the observation of \textbf{W})\textbf{. }The "Wiener filter" $\mathcal{N%
}$ with input \textbf{W, }output $\widetilde{\mathbf{U}}$ and complex gain $%
\eta \left( \omega \right) ,$ is classically defined by%
\begin{equation*}
\eta =\left[ \frac{s_{U}}{s_{W}}\right] =\frac{\left\vert s_{VU}\right\vert
^{2}}{s_{U}s_{V}}
\end{equation*}%
$\widetilde{U}\left( t\right) $ is an estimator of $U\left( t\right) $ based
on the observation of \textbf{V,} with the mean-square error%
\begin{equation*}
\begin{array}{c}
\text{E}\left[ \left\vert U\left( t\right) -\widetilde{U}\left( t\right)
\right\vert ^{2}\right] = \\
\int_{-\infty }^{\infty }\left[ s_{U}\left( 1-\frac{\left\vert
s_{VU}\right\vert ^{2}}{s_{U}s_{V}}\right) \right] \left( \omega \right)
d\omega .%
\end{array}%
\end{equation*}%
The normalized error is%
\begin{equation*}
\frac{1}{\sigma _{U}^{2}}\text{E}\left[ \left\vert U\left( t\right) -%
\widetilde{U}\left( t\right) \right\vert ^{2}\right] =1-\rho _{VU}.
\end{equation*}%
This last equality links the index of coherence $\rho _{VU}$ with the
relative error in the mean-square estimation of the process \textbf{U }from
the observation of the process \textbf{V}. The errorless reconstruction
corresponds to $\rho _{VU}=1,$ and the worse one to $\rho _{VU}=0,$ as
expected$.$

\subsection{Another symmetric index of coherence}

The "spectral degree of coherence at the frequency $f$" of a scalar field is
currently defined by \cite{Wolf2}, \cite{Mand} p. 170,
\begin{equation*}
\mu _{UV}^{0}\left( \omega \right) =\left[ \frac{\left\vert
s_{UV}\right\vert ^{2}}{s_{U}s_{V}}\right] \left( \omega \right)
\end{equation*}%
which depends on the frequency $f=\omega /2\pi .$ The quantity which is
sought herein is a constant. The usual method for reaching this aim is an
integration on the frequency axis. But nothing can assert that%
\begin{equation*}
\mu _{UV}^{1}=\int_{-\infty }^{\infty }\mu _{UV}^{0}\left( \omega \right)
d\omega
\end{equation*}%
verifies the conditions verified by $\rho _{UV}.$ As an example, $\mu
_{UV}^{1}=\infty $ in example 1 of section 2.4.1 with
\begin{equation*}
s_{U}\left( \omega \right) =e^{-\left\vert \omega \right\vert },s_{N}\left(
\omega \right) =e^{-2\left\vert \omega \right\vert }.
\end{equation*}%
For an index which has to characterize a global behavior, the places which
hold a larger power have to be emphazised. For instance, if we weight the
integral which defines $\mu _{UV}^{1}$ by $s_{V}/\sigma _{V}^{2}$, we obtain
the finite index defined in $\left( 8\right) $%
\begin{equation*}
\rho _{UV}=\frac{1}{\sigma _{V}^{2}}\int_{-\infty }^{\infty }\left[ \mu
_{UV}^{0}s_{V}\right] \left( \omega \right) d\omega
\end{equation*}%
but this index is not symmetric ($\rho _{UV}\neq \rho _{VU}$ most of the
time). The symmetry condition is verified by%
\begin{equation*}
\int_{-\infty }^{\infty }\left[ \frac{\left\vert s_{UV}\right\vert ^{2}}{%
\sqrt{s_{U}s_{V}}}\right] \left( \omega \right) d\omega
\end{equation*}%
which is a finite quantity because%
\begin{equation*}
\int_{-\infty }^{\infty }\left[ \sqrt{s_{U}s_{V}}\right] \left( \omega
\right) d\omega <\infty
\end{equation*}%
from the Schwarz inequality. A normalization leads to a new symmetric index
of coherence%
\begin{equation}
\mu _{UV}=\int_{-\infty }^{\infty }\left[ \frac{\left\vert s_{UV}\right\vert
^{2}}{\sqrt{s_{U}s_{V}}}\right] \left( \omega \right) d\omega /\int_{-\infty
}^{\infty }\left[ \sqrt{s_{U}s_{V}}\right] \left( \omega \right) d\omega
\end{equation}%
because $\mu _{UV}$ is a real and positive quantity such that $0\leq \mu
_{UV}\leq 1,$ $\mu _{UV}=0$ if and only if $s_{UV}=0$ ($\mathbf{U}$ and $%
\mathbf{V}$ are uncoherent)$,$ and $\mu _{UV}=1$ if and only if $\mathbf{U}$
and $\mathbf{V}$ are coherent. When $\mu _{UV}$ is different from 0 and 1,
it is also true for $\rho _{UV}$ and $\rho _{VU}.$ The values of these last
quantities are linked to some information held by the process \textbf{U }%
about $V\left( t\right) $ (or the converse) through mean-square estimations.
We do not have such a meaning for $\mu _{UV},$ which is only built to
fulfill some mathematical conditions of normalization and symmetry.

\subsection{Examples}

\subsubsection{Example 1}\label{sec:example1}

\bigskip The simplest model of transmission verifies%
\begin{equation*}
\mathbf{V=U+N}
\end{equation*}%
where \textbf{N }is a "noise" uncorrelated with the "signal" \textbf{U}.
About the decomposition $\left( 6\right) ,$ we find (in accordance with
intuition)%
\begin{equation*}
\mathbf{U=A,N=B}
\end{equation*}%
which leads to the index of coherence%
\begin{equation*}
\rho _{UV}=\frac{\sigma _{U}^{2}}{\sigma _{U}^{2}+\sigma _{N}^{2}}.
\end{equation*}%
The converse is different. $\rho _{VU}$ is obtained from the equations (we
rewrite $\left( 5\right) ,\left( 6\right) ,\left( 8\right) $ inverting
\textbf{U }and \textbf{V})%
\begin{equation*}
\left\{
\begin{array}{c}
\phi =\frac{s_{UV}}{s_{V}}=\frac{s_{U}}{s_{U}+s_{N}} \\
\begin{array}{c}
A\left( t\right) =\mathcal{F}\left[ \mathbf{V}\right] \left( t\right) \\
U\left( t\right) =A\left( t\right) +B\left( t\right) .%
\end{array}%
\end{array}%
\right.
\end{equation*}%
which leads to (using $\left( 8\right) )$%
\begin{equation*}
\rho _{VU}=\frac{1}{\sigma _{U}^{2}}\int_{-\infty }^{\infty }\left[ \frac{%
s_{U}^{2}}{s_{U}+s_{N}}\right] \left( \omega \right) d\omega .
\end{equation*}%
In both situations, we find 1 and 0 as limits following the "ratio" between
the "signal" and the "noise". Between these limits, values of $\rho _{UV}$
and $\rho _{VU}$ are generally different. As an example, let assume that ($%
a>0)$
\begin{equation*}
\left\{
\begin{array}{c}
s_{N}\left( \omega \right) =1-\left\vert \omega \right\vert ,\omega \in
\left( -1,1\right) \text{ and 0 elsewhere} \\
s_{U}\left( \omega \right) =a,\omega \in \left( -1,1\right) \text{ and 0
elsewhere}%
\end{array}%
\right.
\end{equation*}%
which yields
\begin{equation*}
\left\{
\begin{array}{c}
\rho _{UV}=\frac{2a}{1+2a},\rho _{VU}=a\ln \left( 1+\frac{1}{a}\right) \\
\mu _{UV}=3a\frac{\sqrt{1+a}-\sqrt{a}}{\left( 1+a\right) ^{3/2}-a^{3/2}}.%
\end{array}%
\right.
\end{equation*}%
Figure 1 compares the indices as a function of $a.$ The three curves verify
the limit conditions (0 for $a=0$ and 1 for $a=\infty ),$ and are not very
different for other values of $a.$

\begin{figure}[htb]
    \centering
    \centerline{\includegraphics[width=7.5cm]{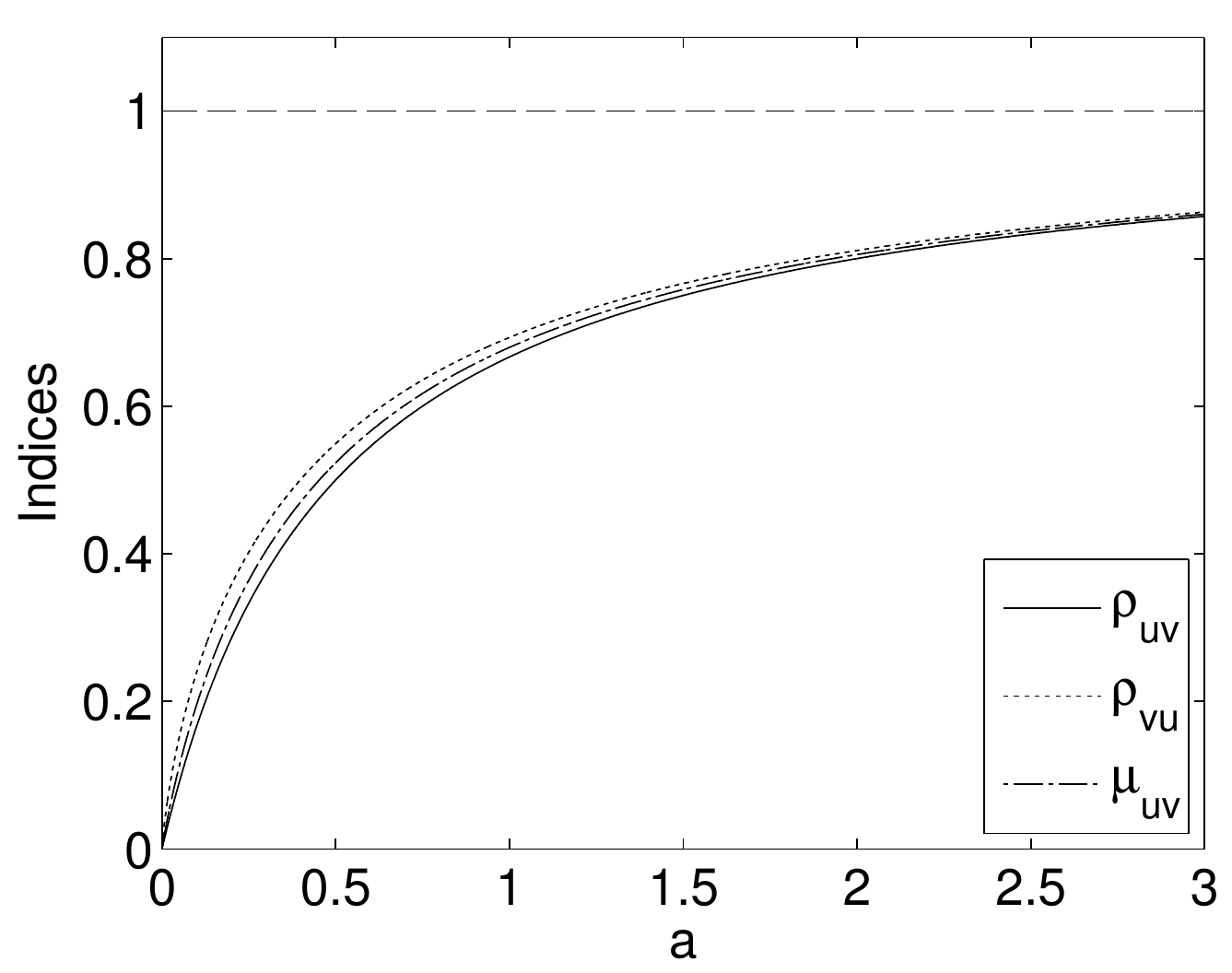}}
    \caption{Example 1 (section \ref{sec:example1}), $\rho_{uv}$, $\rho_{vu}$, $\mu_{uv}$ versus $a$.}
    \label{fig:fig1}
\end{figure}

\subsubsection{Example 2}\label{sec:example2}

Let \textbf{X }be\ a real process independent of \textbf{U }with
characteristic functions (in the probability calculus sense)%
\begin{equation}
\left\{
\begin{array}{c}
\alpha \left( \omega \right) =\text{E}\left[ e^{-i\omega X\left( t\right) }%
\right] \\
\beta \left( \tau ,\omega \right) =\text{E}\left[ e^{-i\omega (X\left(
t\right) -X\left( t-\tau \right) }\right]%
\end{array}%
\right.
\end{equation}%
independent of $t,$ which implies that \textbf{X }is stationary in a sense
stronger than the usual second order one. We define \textbf{V }by%
\begin{equation*}
V\left( t\right) =U\left( t-X\left( t\right) \right) .
\end{equation*}%
$X\left( t\right) $ models a random propagation time which can take into
account variations of \ the refraction index or other hazards \cite{Laca5},
\cite{Laca6}. Easy computations lead to%
\begin{equation}
\left\{
\begin{array}{c}
s_{UV}=\alpha ^{\ast }s_{U}\Longrightarrow \phi =\alpha \\
\rho _{UV}=\frac{1}{\sigma _{U}^{2}}\int_{-\infty }^{\infty }\left[
\left\vert \alpha \right\vert ^{2}s_{U}\right] \left( \omega \right) d\omega
.%
\end{array}%
\right.
\end{equation}%
Moreover,
\begin{equation*}
K_{V}\left( \tau \right) =\int_{-\infty }^{\infty }e^{i\omega \tau }\beta
\left( \tau ,\omega \right) s_{U}\left( \omega \right) d\omega
\end{equation*}%
which allows the determination of $s_{V}\left( \omega \right) $ from a
Fourier transform$.$ Let assume that
\begin{equation*}
s_{U}\left( \omega \right) =\frac{1}{2\pi },\omega \in \left( -\pi ,\pi
\right) \text{ and 0 elsewhere}
\end{equation*}%
and that \textbf{X }is a telegraph signal with values $\pm a$ and parameter $%
\lambda $ (which rules the rate of polarity changes) \cite{Papo}. In this
situation%
\begin{equation*}
\left\{
\begin{array}{c}
\alpha \left( \omega \right) =\cos a\omega \\
\beta \left( \tau ,\omega \right) =\cos ^{2}a\omega +e^{-2\lambda \left\vert
\tau \right\vert }\sin ^{2}a\omega .%
\end{array}%
\right.
\end{equation*}%
From $\left( 8\right) ,$ we obtain%
\begin{equation*}
\rho _{UV}=\frac{1}{2}\left( 1+\frac{\sin 2a\pi }{2a\pi }\right)
\end{equation*}%
which does not depend on $\lambda ,$ and varies from 1 ($a=0)$ to 1/2 ($%
a=\infty ).$ Even for large $a$, $V\left( t+a\right) $ (or $V\left(
t-a\right) )$ provides an estimation of $U\left( t\right) $ which is
errorless approximately half of the time. Moreover, using the convolution
theorem%
\begin{equation*}
\left\{
\begin{array}{c}
s_{V}\left( \omega \right) =\frac{\cos ^{2}a\omega }{2\pi }\Omega \left(
\omega \right) +\frac{1}{\pi ^{2}}\int_{-\pi }^{\pi }\frac{\lambda \sin
^{2}au}{4\lambda ^{2}+\left( \omega -u\right) ^{2}}du \\
\Omega \left( \omega \right) =1,\omega \in \left( -\pi ,\pi \right) \text{
and 0 elsewhere}%
\end{array}%
\right.
\end{equation*}%
which depends on $\lambda ,$ and which never cancels$.$ From $\left(
8\right) $%
\begin{equation*}
\rho _{VU}=\int_{-\pi }^{\pi }\frac{\cos ^{2}a\omega }{4\pi ^{2}s_{V}\left(
\omega \right) }d\omega .
\end{equation*}%
Figure 2 shows variations of $\rho _{VU}$ as function of $\lambda $ (for $%
a=0.1,0.2,0.3,0.5,1,2,4).$ Figure 3 depicts variations of $\rho _{UV}$ and $%
\rho _{VU}$ in function of $a$ for $\lambda =4,8,16.$ As explained above, $%
\rho _{UV}$ and $\rho _{VU}$ are not equal (except for the value 1).

\begin{figure}[htb]
    \centering
    \centerline{\includegraphics[width=7.5cm]{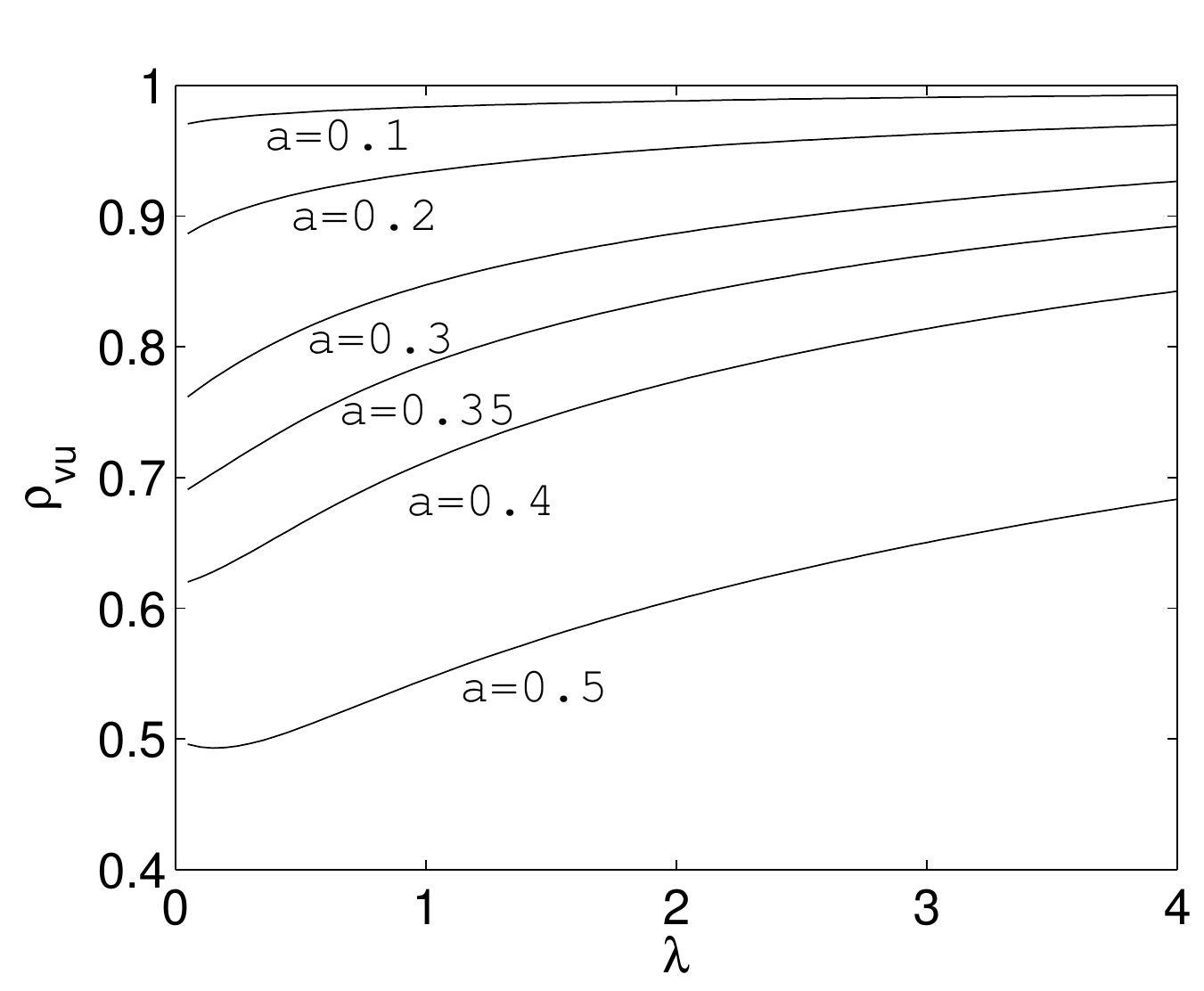}}
    \caption{Example 2 (section \ref{sec:example2}), $\rho_{vu}$ for $a=0.1, 0.2, 0.3, 0.35, 0.4$ and $0.5$ versus $\lambda$.}
    \label{fig:fig2}
\end{figure}

\begin{figure}[htb]
    \centering
    \centerline{\includegraphics[width=7.5cm]{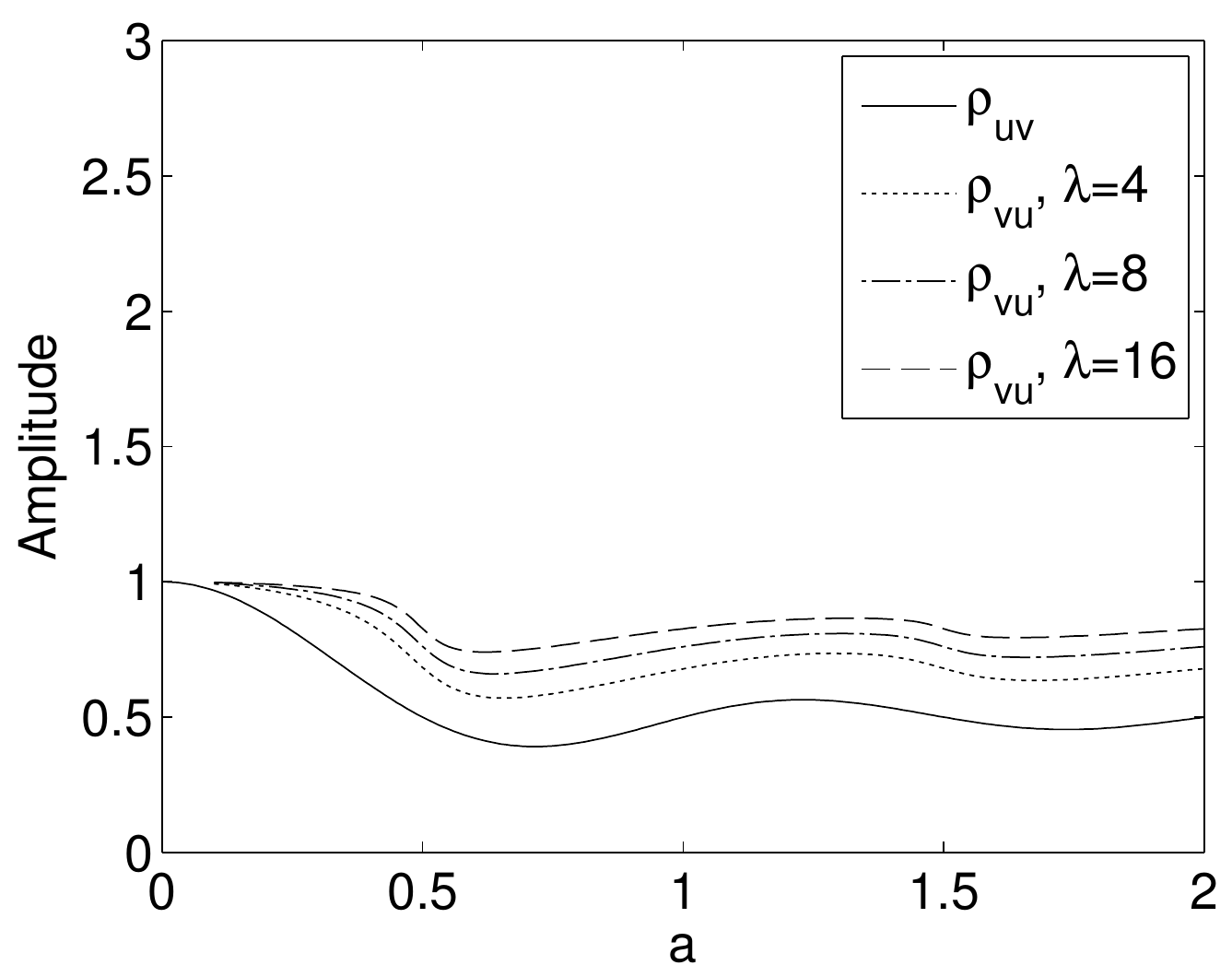}}
    \caption{Example 2 (section \ref{sec:example2}), $\rho_{uv}$ and $\rho_{vu}$ for $\lambda=4, 8$ and $16$.}
    \label{fig:fig3}
\end{figure}

\subsubsection{Example 3}

The lack of symmetry is obvious when spectral supports are not identical.
Let assume that $s_{V}\left( \omega \right) $ does not cancel and that
\textbf{U} is the the result of the low-pass filter with input \textbf{V,}
and complex gain%
\begin{equation*}
\theta \left( \omega \right) =1,\omega \in \left( -b,b\right) \text{ and }0%
\text{ elsewhere.}
\end{equation*}%
In this case, with respect to $\left( 6\right) ,$ we have $\mathbf{A=U}$ and
\textbf{B} is the output of a high-pass filter with input \textbf{V,} and
complex gain%
\begin{equation*}
\theta ^{\prime }\left( \omega \right) =0,\omega \in \left( -b,b\right)
\text{ and }1\text{ elsewhere.}
\end{equation*}%
From $\left( 8\right) $ we deduce%
\begin{equation*}
\rho _{UV}=\frac{1}{\sigma _{V}^{2}}\int_{-b}^{b}s_{V}\left( \omega \right)
d\omega
\end{equation*}%
which verifies $\rho _{UV}<1,$ but we have $\rho _{VU}=1$ because \textbf{U }%
is obtained from \textbf{V} through a LIF.

\subsubsection{Example 4}

1) Let $\mathbf{U}$ be a normalized Gaussian process and \textbf{V }defined
by%
\begin{equation*}
V\left( t\right) =\left\{
\begin{array}{c}
1\text{ when }U\left( t\right) >0 \\
-1\text{ when }U\left( t\right) <0.%
\end{array}%
\right.
\end{equation*}%
Results below are well-known \cite{Papo}%
\begin{equation*}
\left\{
\begin{array}{c}
K_{UV}\left( \tau \right) =\sqrt{\frac{2}{\pi }}K_{U}\left( \tau \right) \\
K_{V}\left( \tau \right) =\frac{2}{\pi }\sin ^{-1}K_{U}\left( \tau \right)%
\end{array}%
\right.
\end{equation*}%
where sin$^{-1}$ is the reciprocal function of the sine function. From $%
\left( 8\right) $ we deduce $\left( K_{U}\left( 0\right) =\sigma
_{U}^{2}=1\right) $%
\begin{equation*}
\rho _{UV}=\frac{2}{\pi },\rho _{VU}=\frac{2}{\pi }\int_{-\infty }^{\infty }%
\left[ \frac{s_{U}^{2}}{s_{V}}\right] \left( \omega \right) d\omega .
\end{equation*}%
Both quantities have no reason to be equal. Moreover, \textbf{U} defines%
\textbf{\ V,} which, in the linear context, is equivalent to \textbf{B=0 }in
$\left( 6\right) $ and $\rho _{UV}=1.$ If this result is false, it is
because the transformation (the sign function) is nonlinear. This function
is not bijective (\textbf{V }gives few informations about \textbf{U}), but
it is not the reason for the small value of $\rho _{UV}.$

2) Let \textbf{V} be defined by
\begin{equation*}
V\left( t\right) =e^{U\left( t\right) }-\sqrt{e}
\end{equation*}%
which is equivalent to%
\begin{equation*}
U\left( t\right) =\ln \left[ V\left( t\right) +\sqrt{e}\right]
\end{equation*}%
provided that $V\left( t\right) >-\sqrt{e}.$ Both processes are coherent in
the sense that \textbf{V }holds informations sufficient for an exact
reconstruction of \textbf{U }(and conversely), but not "linear
informations". Easy computations lead to%
\begin{equation*}
s_{UV}\left( \omega \right) =s_{U}\left( \omega \right) \sqrt{e}.
\end{equation*}%
Consequently, from $\left( 8\right) $%
\begin{equation*}
\rho _{UV}=\frac{1}{e-1}\cong 0.58
\end{equation*}%
and not 1. This counter-example highlights the limits of linear tools. The
property of "coherence" depends on the used mathematical tools. In this
example, both processes are "partially coherent" in the linear framework,
but "coherent" in a wider context.

\subsubsection{Example 5}

It can happen that $0<\rho _{UV}=\rho _{VU}<1.$ It is the case when%
\begin{equation*}
\mathbf{U=X+N,V=X+M}
\end{equation*}%
where \textbf{X }and $\left( \mathbf{N,M}\right) $ are uncorrelated with $%
s_{M}=s_{N}$ (the processes are assumed real and different from \textbf{0}).

\section{Two-dimensional case}

We know that an optical beam is defined by a support, a direction of
propagation and an electrical field orthogonal to this direction. In this
section, we consider two-dimensional processes $\mathbf{U=}\left( \mathbf{U}%
_{x},\mathbf{U}_{y}\right) $ and $\mathbf{V=}\left( \mathbf{V}_{x},\mathbf{V}%
_{y}\right) $ where the components are taken with respect to an orthogonal
system $Oxyz$ where $Oz$ is the direction of propagation. The
four-dimensional process $\left( \mathbf{U}_{x},\mathbf{U}_{y},\mathbf{V}%
_{x},\mathbf{V}_{y}\right) $ is assumed globally stationary with spectral
and cross-spectral densities $s_{U_{x}},...,s_{V_{y}},s_{U_{x}U_{y}},...$
and identical spectral supports. We have to define an "index of coherence"
between 0 and 1, which measures the proximity of the two-dimensional fields
\textbf{U }and \textbf{V},\textbf{\ }and which does not depend on the system
of coordinates $Oxy$. As done previously, we will deduce from one particular
index a natural family of available indices of coherence.

\subsection{A definition of the index of coherence}

As in section 2, we look for the part of \textbf{V=}$\left( \mathbf{V}_{x},%
\mathbf{V}_{y}\right) $\textbf{\ }which is explained by \textbf{U=}$\left(
\mathbf{U}_{x},\mathbf{U}_{y}\right) $. \ As proved in appendix 3, we have
the following decomposition of \textbf{V}%
\begin{equation}
\left\{
\begin{array}{c}
\mathbf{V}_{x}=\mathbf{V}_{x}^{\prime }+\mathbf{V}_{x}^{\prime \prime },%
\mathbf{V}_{y}=\mathbf{V}_{y}^{\prime }+\mathbf{V}_{y}^{\prime \prime } \\
V_{x}^{\prime }\left( t\right) ,V_{y}^{\prime }\left( t\right) \in \mathbf{H}%
_{U_{x}}+\mathbf{H}_{U_{y}} \\
V_{x}^{\prime \prime }\left( t\right) ,V_{y}^{\prime \prime }\left( t\right)
\perp \mathbf{H}_{U_{x}}+\mathbf{H}_{U_{y}}%
\end{array}%
\right.
\end{equation}%
where an element of the set $\mathbf{H}_{U_{x}}+\mathbf{H}_{U_{y}}$ is the
addition of an element of $\mathbf{H}_{U_{x}}$ with an element of $\mathbf{H}%
_{U_{y}}$ (they are the Hilbert spaces spanned by both processes $\mathbf{U}%
_{x},\mathbf{U}_{y}).$ $V_{x}^{\prime }\left( t\right) $ for instance is a
linear combination of the $U_{x}\left( u\right) $ and the $U_{y}\left(
v\right) ,$ and is, in the mean-square sense, the quantity that we can
construct from these r.v. and which is the nearest to $V_{x}\left( t\right)
. $ The parts $\mathbf{V}_{x}^{\prime }$ and $\mathbf{V}_{y}^{\prime }$ hold
informations about \textbf{U}$_{x}$ and \textbf{U}$_{y}$ and not $\mathbf{V}%
_{x}^{\prime \prime }$ and $\mathbf{V}_{y}^{\prime \prime }.$

Appendix 3 shows that the couple of processes $\mathbf{V}^{\prime }=\left(
\mathbf{V}_{x}^{\prime },\mathbf{V}_{y}^{\prime }\right) $ can be retrieved
from five LIF\ $\mathcal{A}_{xx},\mathcal{A}_{xy},\mathcal{M}$,$\mathcal{F}%
_{yx},\mathcal{F}_{yy},$ as depicted in figure 5, with complex gains%
\begin{equation}
\left\{
\begin{array}{c}
\alpha _{xx}=s_{V_{x}U_{x}}/s_{U_{x}},\alpha _{xy}=s_{V_{y}U_{x}}/s_{U_{x}}
\\
\mu =s_{U_{y}U_{x}}/s_{U_{x}} \\
\phi _{yx}=\frac{1}{\Delta }\left[
s_{V_{x}U_{y}}s_{U_{x}}-s_{V_{x}U_{x}}s_{U_{x}U_{y}}\right] \\
\phi _{yy}=\frac{1}{\Delta }\left[
s_{V_{y}U_{y}}s_{U_{x}}-s_{V_{y}U_{x}}s_{U_{x}U_{y}}\right] \\
\Delta =s_{U_{x}}s_{U_{y}}-\left\vert s_{U_{x}U_{y}}\right\vert
^{2}=s_{U_{x}}s_{D} \\
D\left( t\right) =U_{y}\left( t\right) -\mathcal{M}\left[ U_{x}\right]
\left( t\right)%
\end{array}%
\right.
\end{equation}%
where the processes \textbf{U}$_{x}$ and \textbf{D }are orthogonal by
construction. We obtain the decompositions (we omit the variable $t)$%
\begin{equation}
\left\{
\begin{array}{c}
V_{x}=V_{x}^{\prime }+V_{x}^{\prime \prime }=\mathcal{A}_{xx}\left[ \mathbf{U%
}_{x}\right] +\mathcal{F}_{yx}\left[ \mathbf{D}\right] +V_{x}^{\prime \prime
} \\
V_{y}=V_{y}^{\prime }+V_{y}^{\prime \prime }=\mathcal{A}_{xy}\left[ \mathbf{U%
}_{x}\right] +\mathcal{F}_{yy}\left[ \mathbf{D}\right] +V_{y}^{\prime \prime
}%
\end{array}%
\right.
\end{equation}%
where, in both lines, the three terms at right are uncorrelated because, by
construction%
\begin{equation*}
\left\{
\begin{array}{c}
D\left( t\right) \in \mathbf{H}_{U_{x}}+\mathbf{H}_{U_{y}},D\left( t\right)
\perp \mathbf{H}_{U_{x}} \\
V_{x}^{\prime \prime }\left( t\right) ,V_{y}^{\prime \prime }\left( t\right)
\perp \mathbf{H}_{U_{x}}+\mathbf{H}_{U_{y}}.%
\end{array}%
\right.
\end{equation*}%
Consequently, $V_{x}^{\prime }\left( t\right) $ and $V_{y}^{\prime }\left(
t\right) $ are the best (mean-square) estimations of $V_{x}\left( t\right) $
and $V_{y}\left( t\right) $ on $\mathbf{H}_{U_{x}}+\mathbf{H}_{U_{y}}.$ A
reasonable definition of the index of coherence $\rho _{UV}$ between \textbf{%
U }and \textbf{V }is given by%
\begin{equation}
\rho _{UV}=\frac{\text{E}\left[ \left\vert V_{x}^{\prime }\left( t\right)
\right\vert ^{2}+\left\vert V_{y}^{\prime }\left( t\right) \right\vert ^{2}%
\right] }{\text{E}\left[ \left\vert V_{x}\left( t\right) \right\vert
^{2}+\left\vert V_{y}\left( t\right) \right\vert ^{2}\right] }.
\end{equation}%
$\left( \left\vert V_{x}\left( t\right) \right\vert ^{2}+\left\vert
V_{y}\left( t\right) \right\vert ^{2}\right) ^{1/2}$ is the length of $%
V\left( t\right) $ and $\left( \left\vert V_{x}^{\prime }\left( t\right)
\right\vert ^{2}+\left\vert V_{y}^{\prime }\left( t\right) \right\vert
^{2}\right) ^{1/2}$ is the length of its estimation $V^{\prime }\left(
t\right) $. These quantities are independent of the chosen basis and,
obviously, as expected, we have $\rho _{UV}\in \left[ 0,1\right] ,$ with
\begin{equation*}
\left\{
\begin{array}{c}
\rho _{UV}=1\Longleftrightarrow V_{x}\left( t\right) ,V_{y}\left( t\right)
\in \mathbf{H}_{U_{x}}+\mathbf{H}_{U_{y}} \\
\rho _{UV}=0\Longleftrightarrow V_{x}\left( t\right) ,V_{y}\left( t\right)
\perp \mathbf{H}_{U_{x}}+\mathbf{H}_{U_{y}}.%
\end{array}%
\right.
\end{equation*}%
As explained in section 2, the definition has no reason to be symmetric
(generally $\rho _{UV}\neq \rho _{VU}$) and a more general index of
coherence $\rho _{a}$ can be defined:%
\begin{equation}
\rho _{a}=a\rho _{UV}+\left( 1-a\right) \rho _{VU}
\end{equation}%
where $a\in \left[ 0,1\right] .$ Like in section 2, we remark that $\rho
_{1/2}$ is a"symmetric index of coherence", and it is the only one provided
that $\rho _{UV}\neq \rho _{VU}.$

\begin{figure}[htb]
    \centering
    \centerline{\includegraphics[width=7.5cm]{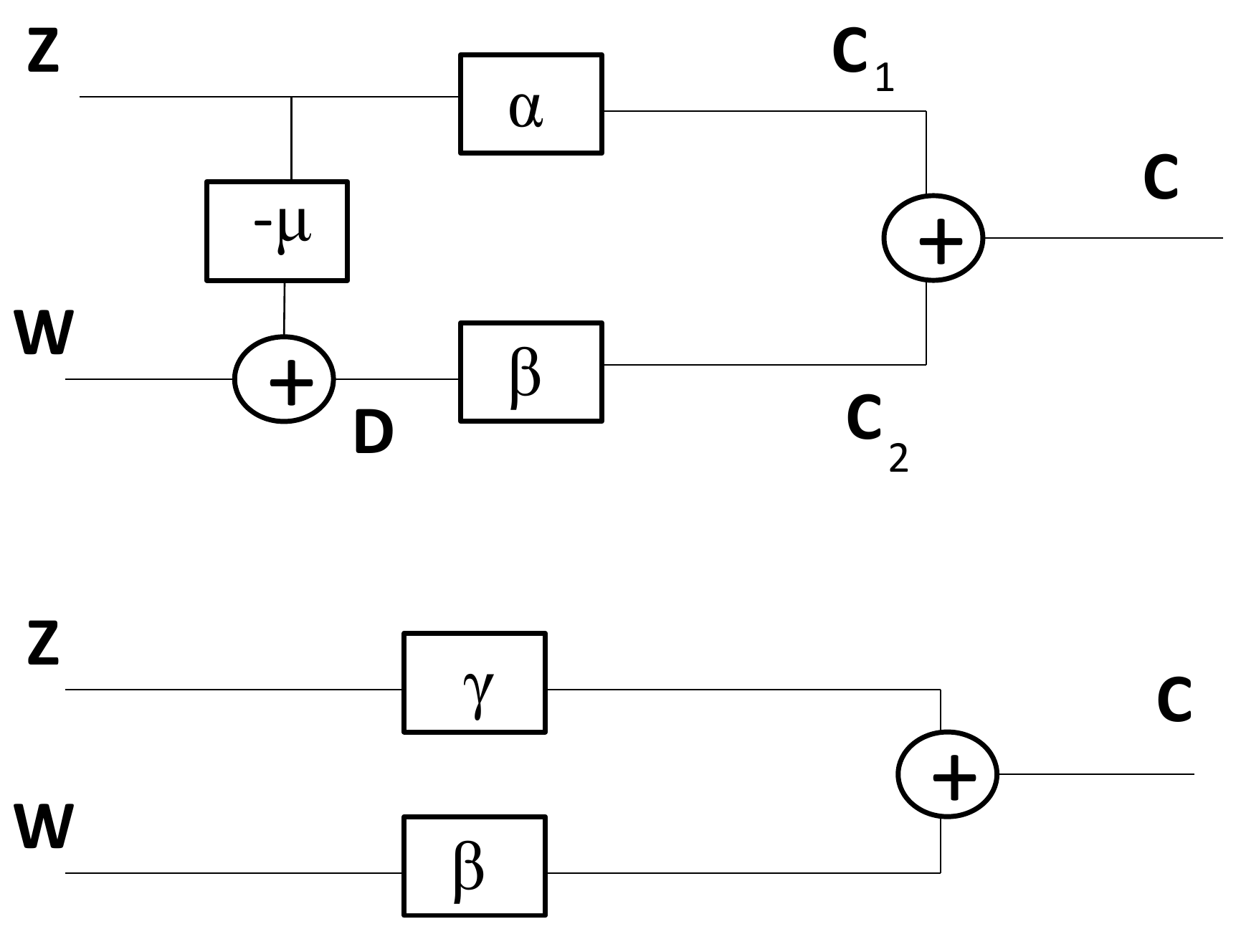}}
    \caption{LIF of complex gains $\alpha, \beta, \gamma, \mu$ defining \textbf{C} from \textbf{Z} and \textbf{W} when $C(t)$ belongs to $\mathbf{H}_{Z}+\mathbf{H}_{W}$.}
    \label{fig:fig4}
\end{figure}

\begin{figure}[htb]
    \centering
    \centerline{\includegraphics[width=7.5cm]{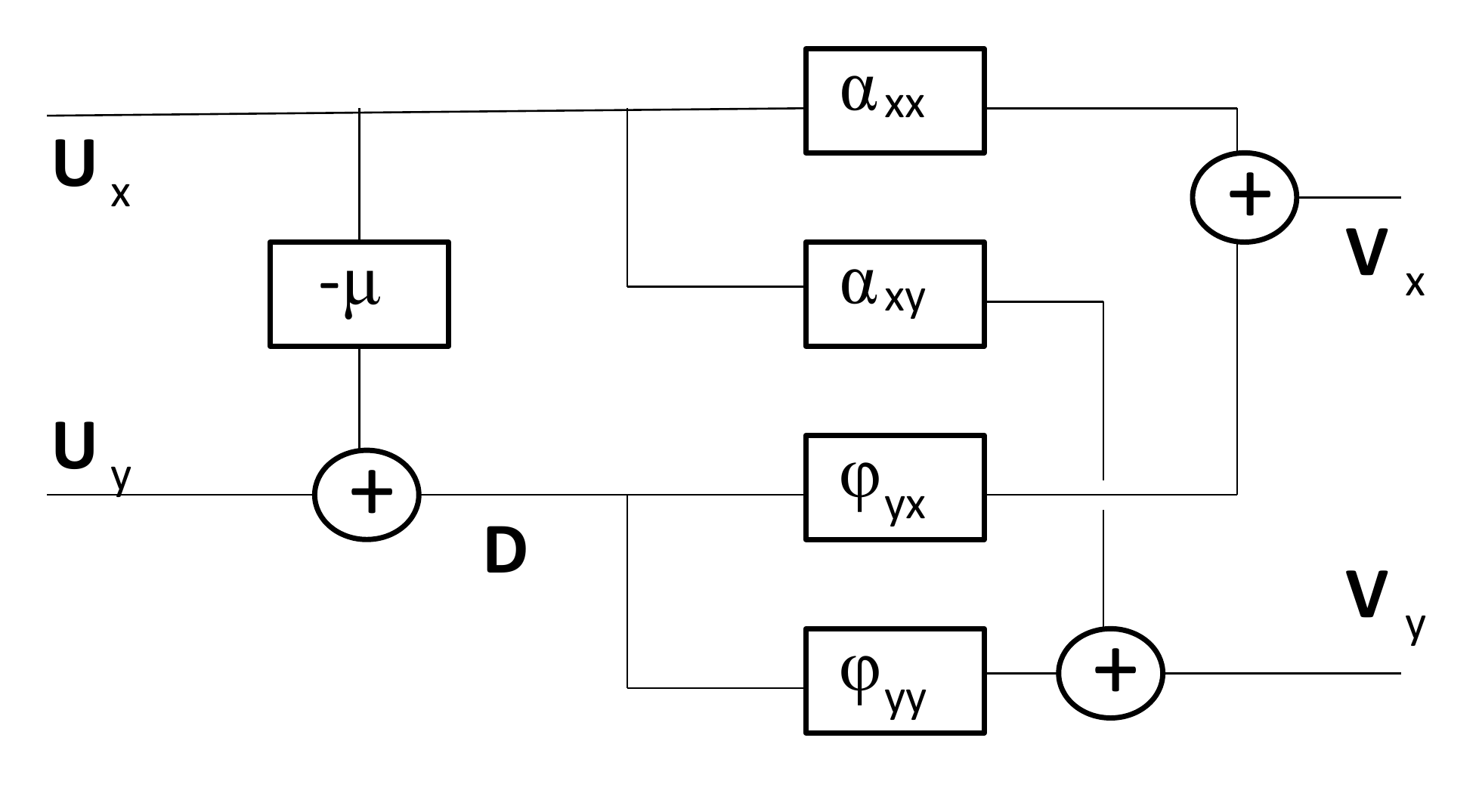}}
    \caption{circuit providing $\mathbf{V}_{x}$ and $\mathbf{V}_{y}$ from $\mathbf{U}_{x}$ and $\mathbf{V}_{y}$ when $V_x(t)$ and $V_y(t)$ belongs to $\mathbf{H}_{U_x}+\mathbf{H}_{V_y}$.}
    \label{fig:fig5}
\end{figure}

\subsection{Estimation}

As shown in the previous section, the best mean-square estimations of $%
V_{x}\left( t\right) $ and $V_{y}\left( t\right) $ from the observation of
the processes \textbf{U}$_{x}$ and \textbf{U}$_{y}$ are $V_{x}^{\prime
}\left( t\right) $ and $V_{y}^{\prime }\left( t\right) $ defined in $\left(
16\right) $ and $\left( 17\right) $ and which are the results of the device
illustrated figure 5.
\begin{equation*}
\left\{
\begin{array}{c}
V_{x}^{\prime }\left( t\right) =\mathcal{A}_{xx}\left[ \mathbf{U}_{x}\right]
\left( t\right) +\mathcal{F}_{yx}\left[ \mathbf{D}\right] \left( t\right) \\
V_{y}^{\prime }\left( t\right) =\mathcal{A}_{xy}\left[ \mathbf{U}_{x}\right]
\left( t\right) +\mathcal{F}_{yy}\left[ \mathbf{D}\right] \left( t\right) .%
\end{array}%
\right.
\end{equation*}%
The estimation errors $\varepsilon _{x}$ and $\varepsilon _{y}$ are usually
defined by
\begin{equation*}
\left\{
\begin{array}{c}
\varepsilon _{x}=\text{E}\left[ \left\vert V_{x}\left( t\right)
-V_{x}^{\prime }\left( t\right) \right\vert ^{2}\right] \\
\varepsilon _{y}=\text{E}\left[ \left\vert V_{y}\left( t\right)
-V_{y}^{\prime }\left( t\right) \right\vert ^{2}\right] \\
\frac{\varepsilon _{x}+\varepsilon _{y}}{\sigma _{V_{x}}^{2}+\sigma
_{V_{y}}^{2}}=1-\rho _{UV}.%
\end{array}%
\right.
\end{equation*}%
Using $\left( 16\right) $ we obtain%
\begin{equation*}
\varepsilon _{x}=\int_{-\infty }^{\infty }\left[ s_{V_{x}}-\frac{\left\vert
s_{V_{x}U_{x}}\right\vert ^{2}}{s_{U_{x}}}-\frac{\Delta \left\vert \phi
_{yx}\right\vert ^{2}}{s_{U_{x}}}\right] \left( \omega \right) d\omega
\end{equation*}%
\begin{equation}
\varepsilon _{y}=\int_{-\infty }^{\infty }\left[ s_{V_{y}}-\frac{\left\vert
s_{VyU_{x}}\right\vert ^{2}}{s_{U_{x}}}-\frac{\Delta \left\vert \phi
_{yy}\right\vert ^{2}}{s_{U_{x}}}\right] \left( \omega \right) d\omega
\end{equation}%
which leads to the formula%
\begin{equation}
\begin{array}{c}
\left( \sigma _{V_{x}}^{2}+\sigma _{V_{y}}^{2}\right) \rho _{UV}= \\
\int_{-\infty }^{\infty }\left[ \frac{\left\vert s_{V_{x}U_{x}}\right\vert
^{2}}{s_{U_{x}}}+\frac{\Delta \left\vert \phi _{yx}\right\vert ^{2}}{%
s_{U_{x}}}\right] \left( \omega \right) d\omega + \\
\int_{-\infty }^{\infty }\left[ \frac{\left\vert s_{VyU_{x}}\right\vert ^{2}%
}{s_{U_{x}}}+\frac{\Delta \left\vert \phi _{yy}\right\vert ^{2}}{s_{U_{x}}}%
\right] \left( \omega \right) d\omega .%
\end{array}%
\end{equation}%
Other formulas are available, replacing respectively in $\left( 16\right) $ $%
U_{x},\phi _{yx},\phi _{yy}$ by $U_{y},\phi _{xy},\phi _{xx}$ defined by%
\begin{equation*}
\left\{
\begin{array}{c}
\phi _{xy}=\frac{1}{\Delta }\left[
s_{V_{y}U_{x}}s_{U_{y}}-s_{V_{y}U_{y}}s_{U_{y}U_{x}}\right] \\
\phi _{xx}=\frac{1}{\Delta }\left[
s_{V_{x}U_{x}}s_{U_{y}}-s_{V_{x}U_{y}}s_{U_{y}U_{x}}\right] .%
\end{array}%
\right.
\end{equation*}

\subsection{Remark}

In modern optics, the "spectral degree of coherence " $%
{{}^\circ}%
c\left( \omega \right) $ is defined by \cite{Wolf2}, \cite{Wolf}%
\begin{equation}
{{}^\circ}%
c\left( \omega \right) =\left[ \frac{s_{U_{x}V_{x}}+s_{U_{y}V_{y}}}{\sqrt{%
\left( s_{U_{x}}+s_{U_{y}}\right) \left( s_{V_{x}}+s_{V_{y}}\right) }}\right]
\left( \omega \right) .
\end{equation}%
$%
{{}^\circ}%
c\left( \omega \right) $ is a complex quantity such that $0\leq \left\vert
{{}^\circ}%
c\left( \omega \right) \right\vert \leq 1.$ The maximum value is obtained
only when%
\begin{equation}
\left\{
\begin{array}{c}
s_{U_{x}V_{x}}\left( \omega \right) =\left[ \sqrt{s_{U_{x}}s_{V_{x}}}\right]
\left( \omega \right) \\
s_{U_{y}V_{y}}\left( \omega \right) =\left[ \sqrt{s_{U_{y}}s_{V_{y}}}\right]
\left( \omega \right) \\
\left[ s_{U_{x}}/s_{U_{y}}\right] \left( \omega \right) =\left[
s_{V_{x}}/s_{V_{y}}\right] \left( \omega \right)%
\end{array}%
\right.
\end{equation}%
which is a condition which separates the coordinates. Now, let assume that
\begin{equation*}
U_{x}\left( t\right) =V_{x}\left( t\right) ,U_{y}\left( t\right)
=3V_{y}\left( t\right) .
\end{equation*}%
We verify that $%
{{}^\circ}%
c\left( \omega \right) <1.$ $\mathbf{U}$ and $\mathbf{V}$ are "coherent" in
the sense where either of them defines the other (and by linear operations
which can be infered). We find the same result when%
\begin{equation*}
U_{x}\left( t\right) =V_{y}\left( t\right) ,U_{y}\left( t\right)
=3V_{x}\left( t\right)
\end{equation*}%
for $\omega \neq 2k\pi ,k\in \mathbb{Z}$, though $\mathbf{U}$ and $\mathbf{V}
$ are still "coherent". We have the same drawback for instance when%
\begin{equation*}
U_{x}\left( t\right) =V_{y}\left( t\right) ,U_{y}\left( t\right)
=-V_{x}\left( t\right)
\end{equation*}%
which can correspond to some rotation of a beam. In this case, we have $%
{{}^\circ}%
c\left( \omega \right) =0$ when the processes \textbf{U}$_{x}$ and \textbf{U}%
$_{y}$ are uncorrelated, thought \textbf{U }defines \textbf{V} perfectly%
\textbf{. }We see through these simple examples that the notion of
"coherence" that is used in this paper is different from the notion defined
by $\left( 22\right) .$

\subsection{Examples}

\subsubsection{Example 1}

Let consider the simple model%
\begin{equation*}
\mathbf{V=U+N}
\end{equation*}%
where \textbf{N=}$\left( \mathbf{N}_{x},\mathbf{N}_{y}\right) $ models an
unpolarized beam\textbf{, }which means that
\begin{equation*}
s_{N_{x}}=s_{N_{y}},s_{N_{x}N_{y}}=0
\end{equation*}%
in any orthonormal basis \cite{Laca4}, \cite{Laca2}. If \textbf{N }is
uncorrelated with \textbf{U, }we have, with respect to $\left( 15\right) $%
\begin{equation*}
\mathbf{V}^{\prime }=\mathbf{U,V}^{\prime \prime }=\mathbf{N}
\end{equation*}%
Consequently (with $\sigma _{N_{x}}=\sigma _{N_{y}}=\sigma _{N})$%
\begin{equation*}
\rho _{UV}=\frac{\sigma _{U_{x}}^{2}+\sigma _{U_{y}}^{2}}{\sigma
_{U_{x}}^{2}+\sigma _{U_{y}}^{2}+2\sigma _{N}^{2}}.
\end{equation*}%
$\rho _{UV}$ decreases from 1 to 0 when $\sigma _{N}$ increases from 0 to $%
\infty ,$ as expected.

When considering%
\begin{equation*}
\mathbf{U=V-N}
\end{equation*}%
we no longer have%
\begin{equation*}
\mathbf{U}^{\prime }=\mathbf{V,U}^{\prime \prime }=-\mathbf{N}
\end{equation*}%
because, for instance%
\begin{equation*}
\text{E}\left[ N_{x}\left( t\right) V_{x}^{\ast }\left( u\right) \right] =%
\text{E}\left[ N_{x}\left( t\right) N_{x}^{\ast }\left( u\right) \right]
\end{equation*}%
has no reason to cancel, and then we do not have
\begin{equation*}
\mathbf{N}_{x},\mathbf{N}_{y}\bot \mathbf{H}_{V_{x}}+\mathbf{H}_{V_{y}}.
\end{equation*}%
The calculus of $\rho _{VU}$ is tedious. We obtain, from $\left( 16\right) $%
\begin{equation*}
\left\{
\begin{array}{c}
\sigma _{U_{x}^{\prime }}^{2}=\int_{-\infty }^{\infty }\left[ \frac{s_{N}}{%
\Delta ^{\prime }}\left( \Delta +s_{N}s_{U_{x}}\right) \right] \left( \omega
\right) d\omega \\
\Delta ^{\prime }=s_{V_{x}}s_{V_{y}}-\left\vert s_{V_{x}V_{y}}\right\vert
^{2} \\
\Delta =s_{U_{x}}s_{U_{y}}-\left\vert s_{U_{x}U_{y}}\right\vert ^{2}%
\end{array}%
\right.
\end{equation*}%
and $\sigma _{U_{y}^{\prime }}^{2}$ by symmetry.

\subsubsection{Example 2}

We consider the model%
\begin{equation*}
\left\{
\begin{array}{c}
V_{x}\left( t\right) =U_{x}\left( t-X\left( t\right) \right) \\
V_{y}\left( t\right) =U_{y}\left( t-X\left( t\right) \right)%
\end{array}%
\right.
\end{equation*}%
where \textbf{X} is defined section 2.4.2 by $\left( 13\right) $. This means
that the propagation is delayed by a quantity which is random and identical
for both components. The processes \textbf{V}$_{x}$ and \textbf{V}$_{y}$ can
be decomposed following the sums \cite{Laca5}%
\begin{equation}
\left\{
\begin{array}{c}
V_{x}\left( t\right) =G_{x}\left( t\right) +Y_{x}\left( t\right) \\
V_{y}\left( t\right) =G_{y}\left( t\right) +Y_{y}\left( t\right)%
\end{array}%
\right.
\end{equation}%
where \textbf{G}$_{x}$ and \textbf{G}$_{y}$ are the outputs of LIF with
respective inputs \textbf{U}$_{x}$ and \textbf{U}$_{y}$ and the same complex
gain $\alpha \left( \omega \right) .$ \textbf{Y}$_{x}$ is uncorrelated
together with \textbf{G}$_{x}$ and \textbf{G}$_{y}.$ The same property is
true for \textbf{Y}$_{y}.$ Consequently,%
\begin{equation}
\left\{
\begin{array}{c}
G_{x}\left( t\right) ,G_{y}\left( t\right) \in \mathbf{H}_{U_{x}}+\mathbf{H}%
_{U_{y}} \\
Y_{x}\left( t\right) ,Y_{y}\left( t\right) \perp \mathbf{H}_{U_{x}}+\mathbf{H%
}_{U_{y}}.%
\end{array}%
\right.
\end{equation}%
The power spectra $s_{Y_{x}}$ and $s_{Y_{y}}$ are different except when $%
s_{U_{x}}=s_{U_{y}}.$ We find:%
\begin{equation}
\rho _{UV}=\frac{1}{\sigma _{U_{x}}^{2}+\sigma _{U_{y}}^{2}}\int_{-\infty
}^{\infty }\left[ \left\vert \alpha \right\vert ^{2}\left(
s_{U_{x}}+s_{U_{y}}\right) \right] \left( \omega \right) d\omega .
\end{equation}%
This result is consistent with intuition. For instance, if $\alpha \left(
\omega \right) =$sinc$\left[ \theta \omega \right] ,$ characteristic
function of a r.v. uniformly distributed on $\left( -\theta ,\theta \right)
, $ the coherence is strong for small $\theta ,$ i.e. for small variations
of the propagation time, and the coherence will be weak for large deviations
of the propagation time (and then for large $\theta ).$ Computations are
harder for $\rho _{VU},$ but are possible, knowing the Fourier transforms of
\begin{equation*}
\left\{
\begin{array}{c}
K_{V_{x}}\left( \tau \right) =\int_{-\infty }^{\infty }e^{i\omega \tau
}\beta \left( \tau ,\omega \right) s_{U_{x}}\left( \omega \right) d\omega \\
K_{V_{y}}\left( \tau \right) =\int_{-\infty }^{\infty }e^{i\omega \tau
}\beta \left( \tau ,\omega \right) s_{U_{y}}\left( \omega \right) d\omega \\
K_{V_{x}V_{y}}\left( \tau \right) =\int_{-\infty }^{\infty }e^{i\omega \tau
}\beta \left( \tau ,\omega \right) s_{U_{x}U_{y}}\left( \omega \right)
d\omega \\
K_{U_{x}V_{x}}\left( \tau \right) =\int_{-\infty }^{\infty }e^{i\omega \tau
} \left[ \alpha ^{\ast }s_{U_{x}}\right] \left( \omega \right) d\omega \\
K_{U_{x}V_{y}}\left( \tau \right) =\int_{-\infty }^{\infty }e^{i\omega \tau
} \left[ \alpha ^{\ast }s_{U_{x}U_{y}}\right] \left( \omega \right) d\omega .%
\end{array}%
\right.
\end{equation*}

\subsubsection{Example 3}

When \textbf{H}$_{V_{x}}+$\textbf{H}$_{V_{y}}$ is included in \textbf{H}$%
_{U_{x}}+$\textbf{H}$_{U_{y}}$, it is clear that $V_{x}\left( t\right) $ and
$V_{y}\left( t\right) $ can be retrieved from $\left( \mathbf{U}_{x},\mathbf{%
U}_{y}\right) $ which is equivalent to $\rho _{UV}=1.$ For instance, it is
the case when%
\begin{equation}
\left\{
\begin{array}{c}
V_{x}\left( t\right) =\mathcal{D}\left[ U_{x}\right] \left( t\right) +%
\mathcal{E}\left[ U_{y}\right] \left( t\right) \\
V_{y}\left( t\right) =\mathcal{F}\left[ U_{x}\right] \left( t\right) -%
\mathcal{G}\left[ U_{y}\right] \left( t\right)%
\end{array}%
\right.
\end{equation}%
where $\mathcal{D}$, $\mathcal{E}$, $\mathcal{F}$, $\mathcal{G}$ are
well-defined LIF. Conversely, $\rho _{VU}=1$ if and only when the linear
system $\left( 27\right) $ can be inverted which is not always possible. For
instance, let assume that the four filters are bandpass on $\left(
-a,a\right) .$ We have (with notations similar to section 3.1)%
\begin{equation*}
\left\{
\begin{array}{c}
U_{x}^{\prime }\left( t\right) =\frac{1}{2}\left[ V_{x}+V_{y}\right] \left(
t\right) \\
U_{y}^{\prime }\left( t\right) =\frac{1}{2}\left[ V_{x}-V_{y}\right] \left(
t\right)%
\end{array}%
\right.
\end{equation*}%
which leads, by using $\left( 18\right) ,$ to%
\begin{equation*}
\rho _{VU}=\int_{-a}^{a}\frac{\left[ s_{U_{x}}+s_{U_{y}}\right] \left(
\omega \right) }{\sigma _{U_{x}}^{2}+\sigma _{U_{y}}^{2}}d\omega .
\end{equation*}

\subsubsection{Example 4}

We study the model%
\begin{equation*}
\left\{
\begin{array}{c}
U_{x}\left( t\right) =X\left( t\right) +M_{x}\left( t\right) ,U_{y}\left(
t\right) =X\left( t\right) +M_{y}\left( t\right) \\
V_{x}\left( t\right) =Y\left( t\right) +N_{x}\left( t\right) ,V_{y}\left(
t\right) =Y\left( t\right) +N_{y}\left( t\right)%
\end{array}%
\right.
\end{equation*}%
where \textbf{M=}$\left\{ \mathbf{M}_{x},\mathbf{M}_{y}\right\} ,$\textbf{N=}%
$\left\{ \mathbf{N}_{x},\mathbf{N}_{y}\right\} $ are unpolarized (see
section 3.4.1) and uncorrelated between them and with $\left( \mathbf{X,Y}%
\right) $. The power spectral densities are $s_{X},s_{Y},s_{M}$ and $s_{N}.$
We find, using $\left( 16\right) $ and $\left( 20\right) $%
\begin{equation*}
\begin{array}{c}
\rho _{UV}=\frac{2}{\sigma _{Y}^{2}+\sigma _{N}^{2}}\int_{-\infty }^{\infty }%
\left[ \frac{\left\vert s_{YX}\right\vert ^{2}}{2s_{X}+s_{M}}\right] \left(
\omega \right) d\omega \\
\rho _{VU}=\frac{2}{\sigma _{X}^{2}+\sigma _{M}^{2}}\int_{-\infty }^{\infty }%
\left[ \frac{\left\vert s_{YX}\right\vert ^{2}}{2s_{Y}+s_{N}}\right] \left(
\omega \right) d\omega%
\end{array}%
\end{equation*}%
The result verifies $\left( 8\right) ,$ when $\sigma _{N}=\sigma _{M}=0.$ We
obtain the same $\rho _{UV}$ with%
\begin{equation*}
\left\{
\begin{array}{c}
U_{x}\left( t\right) =X\left( t\right) \left( \cos \theta -\sin \theta
\right) +M_{x}\left( t\right) \\
U_{y}\left( t\right) =X\left( t\right) \left( \cos \theta +\sin \theta
\right) +M_{y}\left( t\right) \\
V_{x}\left( t\right) =Y\left( t\right) \left( \cos \theta -\sin \theta
\right) +N_{x}\left( t\right) , \\
V_{y}\left( t\right) =Y\left( t\right) \left( \cos \theta +\sin \theta
\right) +N_{y}\left( t\right)%
\end{array}%
\right.
\end{equation*}%
for any $\theta ,$ following the properties of invariance by rotation.

\subsubsection{Example 5}

Finally, let \textbf{V}$_{x},\mathbf{V}_{y}$ be two real processes, and
\textbf{U}$_{x},\mathbf{U}_{y}$ the corresponding analytic signals \cite%
{Papo}. This means that for instance (the integral is defined in the Cauchy
sense)%
\begin{equation*}
U_{x}\left( t\right) =V_{x}\left( t\right) +i\int_{-\infty }^{\infty }\frac{%
V_{x}\left( u\right) }{\pi \left( t-u\right) }du.
\end{equation*}%
We know that the analytic signal loses the negative part of the power
spectrum and we easily find the formulas%
\begin{equation*}
\left\{
\begin{array}{c}
s_{U_{x}}\left( \omega \right) =4s_{V_{x}}\left( \omega \right)
,s_{U_{y}}\left( \omega \right) =4s_{V_{y}}\left( \omega \right) \\
s_{U_{x}U_{y}}\left( \omega \right) =4s_{V_{x}V_{y}}\left( \omega \right)
,\omega >0%
\end{array}%
\right.
\end{equation*}%
and 0 for $\omega <0.$ Obviously, $\rho _{VU}=1.$ But $V_{x}\left( t\right) $
and $V_{y}\left( t\right) $ are the real parts of $U_{x}\left( t\right) $
and $U_{y}\left( t\right) $ and then the former (real) processes can be
deduced from the latter (complex) processes. Nevertheless, the operation
which transforms a complex function in its real part is not linear, which
explains why $\rho _{UV}<1.$ Actually, we find $\rho _{UV}=1/2$, using $%
\left( 16\right) $ and $\left( 18\right) $. This last example shows the
limitations of linear tools, as explained section 2.4.4.

\section{Conclusion\ }

The coherence of a field can be defined as a measure of the proximity
between some properties measured at two points of the field. If the field is
reduced to only one random variable at each point of the space, a
correlation coefficient depending on coordinates of any couple of points may
be a good measure of coherence, the values 0 and 1 addressing the lack of
dependence and, conversely, the complete dependence. When the field is
characterized by one-dimensional stationary processes (for instance \textbf{%
X }and \textbf{Y }at two points\textbf{)}, the normalized cross-correlation $%
\left( 1\right) $ is a natural measure of dependence of $X\left( t-\tau
\right) $ on $Y\left( t\right) ,$ though the latter may be influenced by the
entire set of the $X\left( u\right) ,u\in \mathbb{R}$, and not only by the
value $X\left( t-\tau \right) $ of \textbf{X }at $t-\tau .$ If the entire
process \textbf{X }is observed, and assuming the stationarity property,
formula $\left( 1\right) $ does not provide the whole available information
held by \textbf{X }about the elements of \textbf{Y}$.$ A positive number
which measures global links between $X\left( t\right) $\textbf{\ }and the
entire process \textbf{Y }(or the converse)\textbf{\ }appears to be a \
better characterization of proximity of both processes\textbf{. }It seems
equivalent looking for a copy $\widetilde{\mathbf{X}}$ of $\mathbf{X}$ from
\textbf{Y }(an estimation of $X\left( t\right) $ on the entire set of the $%
Y\left( u\right) $) and conversely, which is an usual procedure in
communications. Obviously, the idea of coherence is linked to the similarity
between the model and the copy. At the next stage, we compute a distance
(the error) between both and we define an index of coherence normalizing the
latter. We have explained the main drawback of this construction: it leads
to a different index when \textbf{X }and \textbf{Y }are inverted.
Nevertheless, this enables the definition of an available linear family of
indices. We show that other constructions can be achieved which lead to a
symmetric index of coherence. When the field is no longer one-dimensional
but two-dimensional, definitions are generalized. The main idea is
unchanged, which looks for characterizing a kind of distance between Hilbert
spaces respectively spanned by each of two-dimensional processes. The
resulting "index of coherence" is still a number between 0 and 1 as expected
and not some function of time or frequency. Examples are given to cover a
sufficient number of situations, and appendices summarize the main results
of the stationary process theory, and detail laborious calculations.

\section{Appendices}

\subsection{Appendix 1: notations}

1) Let \textbf{U=}$\left\{ U\left( t\right) ,t\in \mathbb{R}\right\} $ be\ a
zero-mean stationary process. Auto-correlation function $K_{U},$ spectral
density $s_{U}$ and total power $\sigma _{U}^{2}$ verify%
\begin{equation*}
\begin{array}{c}
K_{U}\left( \tau \right) =\text{E}\left[ U\left( t\right) U^{\ast }\left(
t-\tau \right) \right] =\int_{-\infty }^{\infty }s_{U}\left( \omega \right)
e^{i\omega \tau }d\omega \\
\sigma _{U}^{2}=K_{U}\left( 0\right) .%
\end{array}%
\end{equation*}%
where E$\left[ ..\right] $ and the superscript $^{\ast }$ stand for the
mathematical expectation (ensemble mean) the complex conjugate.

2) The cross-correlation $K_{UV},$ the cross-spectral density $s_{UV}$
between the processes \textbf{U }and \textbf{V }are defined by%
\begin{equation*}
K_{UV}\left( \tau \right) =\text{E}\left[ U\left( t\right) V^{\ast }\left(
t-\tau \right) \right] =\int_{-\infty }^{\infty }s_{UV}\left( \omega \right)
e^{i\omega \tau }d\omega
\end{equation*}%
when both processes are stationary and have stationary cross-correlations
(equivalently $\left( \mathbf{U,V}\right) $ is stationary). All these
quantities are always assumed regular enough.\

3) \textbf{H}$_{U}$ is the Hilbert space of linear combinations of the $%
U\left( t\right) ,t\in \mathbb{R}.$ This means that (for some $t_{kn}\in
\mathbb{R}$, $a_{kn}\in \mathbb{C}$)%
\begin{equation*}
A\in \mathbf{H}_{U}\Longleftrightarrow A=\lim_{n\rightarrow \infty
}\sum_{k=-n}^{n}a_{kn}U\left( t_{kn}\right)
\end{equation*}%
in the mean-square sense. The scalar product $\left\langle .,.\right\rangle
_{\mathbf{H}_{U}}$ in \textbf{H}$_{U}$ is defined by%
\begin{equation*}
\left\langle A,B\right\rangle _{\mathbf{H}_{U}}=\text{E}\left[ AB^{\ast }%
\right] .
\end{equation*}

4) \textbf{K}$_{s_{U}}$ is the Hilbert space of complex valued functions $f$
such that%
\begin{equation*}
\int_{-\infty }^{\infty }\left[ \left\vert f\right\vert ^{2}s_{U}\right]
\left( \omega \right) d\omega <\infty .
\end{equation*}%
The scalar product $\left\langle .,.\right\rangle _{\mathbf{K}_{s_{U}}}$ is
defined by%
\begin{equation*}
\left\langle f,g\right\rangle _{\mathbf{K}_{s_{U}}}=\int_{-\infty }^{\infty }%
\left[ fg^{\ast }s_{U}\right] \left( \omega \right) d\omega .
\end{equation*}

5) The isometry \textbf{I}$_{U}$ between \textbf{H}$_{U}$ and \textbf{K}$%
_{s_{U}}$is defined from the correspondence%
\begin{equation*}
U\left( t\right) \Longleftrightarrow _{\mathbf{I}_{U}}e^{i\omega t}.
\end{equation*}%
If $A=\lim_{n\rightarrow \infty }\sum_{k=-n}^{n}a_{kn}U\left( t_{kn}\right)
, $ then%
\begin{equation*}
A\Longleftrightarrow _{\mathbf{I}_{U}}\lim_{n\rightarrow \infty
}\sum_{k=-n}^{n}a_{kn}e^{i\omega t_{kn}}.
\end{equation*}%
Moreover, if $A\Longleftrightarrow _{\mathbf{I}_{U}}\alpha
,B\Longleftrightarrow _{\mathbf{I}_{U}}\beta ,$ then%
\begin{equation*}
\text{E}\left[ \left\vert A-B\right\vert ^{2}\right] =\int_{-\infty
}^{\infty }\left[ \left\vert \alpha -\beta \right\vert ^{2}s_{U}\right]
\left( \omega \right) d\omega .
\end{equation*}%
The isometry allows to solve a problem of distance between random variables
(r.v.) using Fourier analysis.

6) The Linear Invariant Filter (LIF) $\mathcal{F}$ with complex gain $\phi ,$
input $U\left( t\right) ,$ output $V\left( t\right) $ is defined by%
\begin{equation*}
V\left( t\right) =\mathcal{F}\left[ \mathbf{U}\right] \left( t\right)
\Longleftrightarrow _{\mathbf{I}_{U}}\phi \left( \omega \right) e^{i\omega
t}.
\end{equation*}%
The impulse response $f$ of $\mathcal{F}$ is defined by (in some sense)%
\begin{equation*}
\phi \left( \omega \right) =\int_{-\infty }^{\infty }f\left( u\right)
e^{-i\omega u}du.
\end{equation*}%
For a regular enough $f\,,$ we have%
\begin{equation*}
\mathcal{F}\left[ \mathbf{U}\right] \left( t\right) =\int_{-\infty }^{\infty
}f\left( u\right) U\left( t-u\right) du.
\end{equation*}%
If $W\left( t\right) =\mathcal{G}\left[ \mathbf{U}\right] \left( t\right) $
is the output of the LIF of complex gain $\gamma ,$ we have%
\begin{equation*}
\begin{array}{c}
\text{E}\left[ V\left( t\right) W^{\ast }\left( t-\tau \right) \right]
=\int_{-\infty }^{\infty }\left[ \phi \gamma ^{\ast }s_{U}\right] \left(
\omega \right) e^{i\omega \tau }d\omega \\
s_{VW}\left( \omega \right) =\left[ \phi \gamma ^{\ast }s_{U}\right] \left(
\omega \right) .%
\end{array}%
\end{equation*}%
This relation is known as the "theorem of interferences".

Though the principles above are very general, we assume that the used
processes have bounded spectral densities. Nevertheless, results in this
paper are true for monochromatic waves, which are approximations of waves
encountered in the real word.

\subsection{Appendix 2}

Let assume that $A\left( t\right) $ is the orthogonal projection of $V\left(
t\right) $ on $\mathbf{H}_{U}:$%
\begin{equation*}
A\left( t\right) =\text{pr}_{\mathbf{H}_{U}}V\left( t\right) .
\end{equation*}%
This means that $V\left( t\right) -A\left( t\right) $ is orthogonal to any $%
U\left( u\right) $ (the r.v. which generate $\mathbf{H}_{U}):$
\begin{equation*}
\text{E}\left[ \left( V\left( t\right) -A\left( t\right) \right) U^{\ast
}\left( u\right) \right] =0
\end{equation*}%
for any $u\in \mathbb{R}$. Equivalently, whatever $u$%
\begin{equation*}
\int_{-\infty }^{\infty }\left( s_{VU}\left( \omega \right) e^{i\omega
\left( t-u\right) }-\left[ \phi _{t}s_{U}\right] \left( \omega \right)
e^{-i\omega u}\right) d\omega =0
\end{equation*}%
when, in the usual isometry \textbf{I}$_{U}$ built from \textbf{H}$_{U},$ we
have the correspondences%
\begin{equation*}
U\left( t\right) \longleftrightarrow _{\mathbf{I}_{U}}e^{i\omega t},A\left(
t\right) \longleftrightarrow _{\mathbf{I}_{U}}\phi _{t}\left( \omega \right)
.
\end{equation*}%
As a consequence of the unicity of the Fourier transform, we deduce%
\begin{equation*}
\phi _{t}\left( \omega \right) =\left[ \frac{s_{VU}}{s_{U}}\right] \left(
\omega \right) e^{i\omega t}
\end{equation*}%
which means that \textbf{A }is the output of the LIF with input \textbf{U }%
and complex gain%
\begin{equation*}
\phi \left( \omega \right) =\left[ \frac{s_{VU}}{s_{U}}\right] \left( \omega
\right) .
\end{equation*}%
In the equality $\left( 6\right) ,$ $B\left( t\right) $ is orthogonal to
\textbf{H}$_{U}$ and then orthogonal together to the $U\left( u\right) $ and
the $A\left( u\right) .$

\subsection{Appendix 3}

We consider three stationary processes \textbf{Z, W, C. }We assume that $%
\left( \mathbf{Z,W}\right) $ is stationary, that \textbf{C} has stationary
correlations with $\left( \mathbf{Z,W}\right) $ and that $C\left( t\right)
\in \mathbf{H}_{Z}+\mathbf{H}_{W}$ (which means that $C\left( t\right) $ is
the result of linear operations from $\mathbf{Z}$ and $\mathbf{W}).$ We have
to justify the drawings of figure 4, where $\mu \left( \omega \right)
,\alpha \left( \omega \right) ,\beta \left( \omega \right) ,\gamma \left(
\omega \right) $ are complex gains of LIF to be characterized. $%
s_{Z},s_{ZW}...$ are the spectral densities and the cross-spectra.

1) Let \textbf{C}$_{1}$ be defined by%
\begin{equation*}
C_{1}\left( t\right) =\text{pr}_{\mathbf{H}_{Z}}C\left( t\right) .
\end{equation*}%
If $C_{1}\left( t\right) \longleftrightarrow _{I_{Z}}\alpha _{t}\left(
\omega \right) ,$ we have (whatever $t,u\in \mathbb{R)}$%
\begin{equation*}
\begin{array}{c}
\text{E}\left[ \left( C\left( t\right) -C_{1}\left( t\right) \right) Z^{\ast
}\left( u\right) \right] =0\mathbb{\Longleftrightarrow } \\
\int_{-\infty }^{\infty }e^{-i\omega u}\left[ e^{i\omega t}s_{CZ}\left(
\omega \right) -\left[ \alpha _{t}s_{Z}\right] \left( \omega \right) \right]
d\omega =0%
\end{array}%
\end{equation*}%
which is equivalent to%
\begin{equation*}
\alpha _{t}\left( \omega \right) =e^{i\omega t}\left[ \frac{s_{CZ}}{s_{Z}}%
\right] \left( \omega \right) .
\end{equation*}%
Consequently, \textbf{C}$_{1}$ is the output of a LIF with complex gain%
\begin{equation}
\alpha \left( \omega \right) =\left[ \frac{s_{CZ}}{s_{Z}}\right] \left(
\omega \right) .
\end{equation}

2) Let \textbf{D} be defined by
\begin{equation*}
D\left( t\right) =W\left( t\right) -\mathcal{M}\left[ \mathbf{Z}\right]
\left( t\right)
\end{equation*}%
where $\mathcal{M}$ is some LIF complex gain $\mu \left( \omega \right) .$
We look for $\mu $ such that $D\left( t\right) \bot \mathbf{H}_{Z}:$%
\begin{equation*}
\begin{array}{c}
\text{E}\left[ \left( W\left( t\right) -\mu \left[ \mathbf{Z}\right] \left(
t\right) \right) Z^{\ast }\left( u\right) \right] =0\Longleftrightarrow \\
\int_{-\infty }^{\infty }e^{i\omega \left( t-u\right) }\left[ s_{WZ}-\mu
s_{Z}\right] \left( \omega \right) d\omega =0%
\end{array}%
\end{equation*}%
which yields%
\begin{equation}
\mu \left( \omega \right) =\left[ \frac{s_{WZ}}{s_{Z}}\right] \left( \omega
\right) .
\end{equation}%
Moreover, we remark that%
\begin{equation*}
\mathbf{H}_{Z}+\mathbf{H}_{W}=\mathbf{H}_{Z}+\mathbf{H}_{D}
\end{equation*}%
because $W\left( t\right) =D\left( t\right) +\mathcal{M}\left[ \mathbf{Z}%
\right] \left( t\right) .$ Also%
\begin{equation}
\begin{array}{c}
s_{D}\left( \omega \right) =\left[ s_{W}-\frac{\left\vert s_{ZW}\right\vert
^{2}}{s_{Z}}\right] \left( \omega \right) \\
s_{DC}\left( \omega \right) =\left[ s_{WC}-\frac{s_{WZ}s_{ZC}}{s_{Z}}\right]
\left( \omega \right) .%
\end{array}%
\end{equation}

3) Let \textbf{C}$_{2}$ defined by
\begin{equation*}
C_{2}\left( t\right) =\text{pr}_{\mathbf{H}_{D}}C\left( t\right) .
\end{equation*}%
By construction, $\mathbf{H}_{Z}$ and $\mathbf{H}_{D}$ are orthogonal and $%
\mathbf{H}_{C}\subset \mathbf{H}_{Z}$ +$\mathbf{H}_{W}$ by hypothesis, which
implies
\begin{equation*}
C\left( t\right) =C_{1}\left( t\right) +C_{2}\left( t\right) .
\end{equation*}%
The problem is to prove that \textbf{C}$_{2}$ is the output of a LIF. If $%
C_{2}\left( t\right) \longleftrightarrow _{I_{D}}\beta _{t}\left( \omega
\right) ,$ we have%
\begin{equation*}
\begin{array}{c}
\text{E}\left[ \left( C\left( t\right) -C_{2}\left( t\right) \right) D^{\ast
}\left( u\right) \right] =0\Longleftrightarrow \\
\int_{-\infty }^{\infty }e^{-i\omega u}\left[ e^{i\omega t}\phi \left(
\omega \right) -\left[ \beta _{t}s_{D}\right] \left( \omega \right) \right]
d\omega =0. \\
\phi \left( \omega \right) =\left[ s_{CW}-\mu ^{\ast }s_{CZ}\right] \left(
\omega \right)%
\end{array}%
\end{equation*}%
Using $\left( 29\right) $ and $\left( 30\right) $ we obtain%
\begin{equation*}
\beta _{t}\left( \omega \right) =e^{i\omega t}\left[ \frac{%
s_{CW}s_{Z}-s_{CZ}s_{ZW}}{s_{W}s_{Z}-\left\vert s_{ZW}\right\vert ^{2}}%
\right] \left( \omega \right)
\end{equation*}%
which proves that \textbf{C}$_{2}$ is the output of a LIF $\mathcal{B}$ with
input \textbf{D} and complex gain%
\begin{equation}
\beta \left( \omega \right) =\left[ \frac{s_{CW}s_{Z}-s_{CZ}s_{ZW}}{%
s_{W}s_{Z}-\left\vert s_{ZW}\right\vert ^{2}}\right] \left( \omega \right) .
\end{equation}%
Figure 4 depicts a symmetric equivalent circuit which highlights the LIF of
complex gain $\gamma \left( \omega \right) $ with%
\begin{equation}
\gamma \left( \omega \right) =\left[ \frac{s_{CZ}s_{W}-s_{CW}s_{WZ}}{%
s_{W}s_{Z}-\left\vert s_{ZW}\right\vert ^{2}}\right] \left( \omega \right) .
\end{equation}%
As a consequence, the power spectrum $s_{C}$ verifies%
\begin{equation*}
\begin{array}{c}
s_{C}\left( \omega \right) = \\
\left[ \frac{\left\vert s_{CZ}\right\vert ^{2}s_{W}+\left\vert
s_{CW}\right\vert ^{2}s_{Z}-2\mathcal{R}\left[ s_{CW}s_{WZ}s_{ZC}\right] }{%
s_{W}s_{Z}-\left\vert s_{WZ}\right\vert ^{2}}\right] \left( \omega \right) .%
\end{array}%
\end{equation*}%
Moreover, the symmetric scheme is unique, provided that the set of $\omega $
such that%
\begin{equation*}
\left\vert s_{ZW}\right\vert \left( \omega \right) \neq \sqrt{s_{Z}s_{W}}%
\left( \omega \right)
\end{equation*}%
has a positive measure.

4) If $\mathbf{C}$ and $\mathbf{C}^{\prime }$ have stationary correlations
with $\left( \mathbf{Z,W}\right) $ and belong to $\mathbf{H}_{Z}+\mathbf{H}%
_{W}$, then $\left( \mathbf{C},\mathbf{C}^{\prime }\right) $ is stationary
and with cross-spectrum%
\begin{equation*}
\left\{
\begin{array}{c}
s_{CC^{\prime }}\left( \omega \right) = \\
\left[ \frac{as_{CZ}-bs_{CW}}{s_{W}s_{Z}-\left\vert s_{WZ}\right\vert ^{2}}%
\right] \left( \omega \right) \\
a\left( \omega \right) =\left[ s_{W}s_{ZC^{\prime }}-s_{ZW}s_{WC^{\prime }}%
\right] \left( \omega \right) \\
b\left( \omega \right) =\left[ s_{Z}s_{WC^{\prime }}-s_{WZ}s_{ZC^{\prime }}%
\right] \left( \omega \right) .%
\end{array}%
\right.
\end{equation*}%
To summarize, if $\mathbf{C}$ and $\mathbf{C}^{\prime }$ are stationarily
correlated with $\mathbf{Z}$ and $\mathbf{W}$ and belong to $\mathbf{H}_{Z}+%
\mathbf{H}_{W}\mathbf{,}$ they are the outputs of a "bi-filter" with
perfectly determined components.

5) When the hypothesis $C\left( t\right) \in \mathbf{H}_{Z}+\mathbf{H}_{W}$
is suppressed, we have the decomposition%
\begin{equation*}
\begin{array}{c}
C\left( t\right) =C^{\prime }\left( t\right) +C^{\prime \prime }\left(
t\right) \\
C^{\prime }\left( t\right) =\text{pr}_{\mathbf{H}_{Z}+\mathbf{H}_{W}}C\left(
t\right) \\
C^{\prime \prime }\left( t\right) \perp \mathbf{H}_{Z}+\mathbf{H}_{W}.%
\end{array}%
\end{equation*}%
$\mathbf{\ C}^{\prime }$ is stationarily correlated with $\left( \mathbf{Z,W}%
\right) $ because%
\begin{equation*}
\text{E}\left[ C^{\prime }\left( t\right) \left[ aZ+bW\right] ^{\ast }\left(
u\right) \right] =\left[ aK_{CZ}+bK_{CW}\right] \left( t-u\right) .
\end{equation*}%
As a consequence, the drawings in figure 4 are available, replacing $\mathbf{%
C}$ by $\mathbf{C}^{\prime }$ as output (but with the same complex gains $%
\mu ,\alpha ,\beta ,\gamma ).$

\end{document}